\documentclass[11pt,a4paper]{article}
\usepackage[margin=2cm]{geometry}
\usepackage[utf8]{inputenc}
\usepackage{authblk}
\usepackage{graphicx}
\usepackage{hyperref}
\usepackage{amsmath}
\usepackage{amssymb}
\usepackage{slashed}
\usepackage{color}
\usepackage{booktabs}
\usepackage[shortlabels]{enumitem}
\usepackage[nameinlink,capitalize]{cleveref}

\newcommand{\uone}{U(1)_\mathrm{FN}}
\newcommand{\vt}{v_{\theta}}
\newcommand{\vs}{v_{\sigma}}
\newcommand{\Lfn}{\Lambda_\mathrm{FN}}
\newcommand{\Lew}{\Lambda_\mathrm{EW}}
\newcommand{\Lsi}{\Lambda_\Sigma}
\newcommand{\mus}{\mu_{\sigma}^2}
\newcommand{\mut}{\mu_{\theta}^2}
\newcommand{\ls}{\lambda_{\sigma}}
\newcommand{\lt}{\lambda_{\theta}}
\newcommand{\lts}{\lambda_{\theta \sigma}}

\title{\textbf{Seesaw neutrino dark matter by freeze-out}}
\author{Carlos Jaramillo\footnote{email: carlos.jaramillo@mpi-hd.mpg.de},\quad  Manfred Lindner, \quad Werner Rodejohann}
\affil{\textit{Max-Planck-Institut f\"ur Kernphysik, Saupfercheckweg 1, 69117 Heidelberg, Germany}}
\date{}

\begin{document}

\maketitle
{\small {\flushleft \url{https://doi.org/10.1088/1475-7516/2021/04/023} \hfill  JCAP04(2021)023}}

\begin{abstract}\noindent
    We investigate whether right-handed neutrinos can play the role of the dark matter of the Universe and be generated by the  freeze-out production mechanism. In the standard picture, the requirement of a long lifetime of the right-handed neutrinos implies a small neutrino Yukawa coupling. As a consequence, they never reach thermal equilibrium, thus prohibiting production by freeze-out.
    We note that this limitation is alleviated if the neutrino Yukawa coupling is large enough in the early Universe to thermalize the sterile neutrinos, and then becomes tiny at a certain moment, which makes them drop out of equilibrium.
    As a concrete example realization of this framework, we consider a Froggatt-Nielsen  model supplemented by an additional scalar field which obeys a global symmetry (not the flavour symmetry). Initially, the vacuum expectation value of the flavon is such, that the effective neutrino Yukawa coupling is large and unsuppressed, keeping them in thermal equilibrium.
    At some point the new scalar also gets a vacuum expectation value that breaks the symmetry. This may occur in such a way that the vev of the flavon is shifted to a new (smaller) value. In that case, the Yukawa coupling is reduced such that the sterile neutrinos are rendered stable on cosmological time scales.  We show that this mechanism works for a wide range of sterile neutrino masses.
\end{abstract}

\section{Introduction}
The identity and the character of the Dark Matter (DM) of the Universe and the origin of the minute mass of neutrinos in the Standard Model of particle physics (SM) are two of the most prominent problems in fundamental physics. Particularly the dark matter problem, independently of its eventual solution, can be considered a hint for the incompleteness of the SM as a theory of nature.

One of the simplest mechanisms that explains not only the origin of the masses of the SM neutrinos, but also offers a natural explanation for their tiny magnitude, is the type I seesaw mechanism \cite{Minkowski:1977sc,Yanagida:1979as,GellMann:1980vs,Mohapatra:1980yp}: 
Introducing three right-handed neutrinos (which are SM singlets) one can write a Yukawa term including those and the three lepton doublets, which after electroweak symmetry breaking (EWSB) generates a Dirac mass for the neutrinos, analogously to the up-quark sector.
Furthermore, a Majorana mass term for the right-handed neutrinos is  allowed, since it does not violate any SM gauge symmetry. 
After diagonalization, the resulting neutrino mass matrix delivers three very light  neutrino-mass eigenstates (mostly composed of the left-handed states) and on the other hand very heavy mass eigenstates largely composed of the right handed neutrinos.

Assuming that the type I seesaw mechanism is indeed responsible for the masses of SM neutrinos, one immediately wonders whether one of the new sterile neutrinos might play the role of the dark matter particle.
A massive neutral SM singlet is in fact the prototype of a weakly interacting massive particle (WIMP), the still most popular candidate for the observationally favored cold dark matter.
The next question would be whether the observed DM density can be generated by thermal freeze-out, i.e.\  whether, after entering thermal equilibrium, its rate with the SM plasma at some stage can no longer compete with the expansion of the Universe. 
This production mechanism is well known and has been thoroughly studied in the context of WIMP dark matter.
Arguably, it is the most attractive and natural production mechanism of dark matter. 
For the right-handed neutrinos to be DM, they have to fulfil the following conditions:
\begin{enumerate}
    \item \textbf{Abundance:} the sterile neutrino has to be produced in the early Universe to such an amount, that it contributes meaningfully to the observed dark matter abundance $\Omega_\mathrm{DM}h^2 \approx 0.12$ \cite{Aghanim:2018eyx}. 
		It does not need to saturate this amount, but it definitely may not surpass it.
    \item \textbf{Longevity:} dark matter must be stable on cosmological time scales. If it is able to decay, then the lifetime of the decay must be comparable or larger than the age of the Universe.
    \item \textbf{Constraints:} Needless to say, any viable dark matter model must comply with any observational or experimental constraints that apply to it.
\end{enumerate}

The first thing that comes to mind in the context of dark matter and right-handed neutrinos is the possibility of keV-scale sterile neutrinos, which are ideal warm dark matter candidates, for reviews see \cite{Adhikari:2016bei,Boyarsky:2018tvu}.
Their production typically works via oscillations with active neutrinos \cite{Dodelson1994}, potentially enhanced by resonances from lepton asymmetries \cite{Shi:1998km}, or with additional input in the form of the decay of new heavier particles \cite{Alanne2018, Kusenko2009}.
The mixing of the particles with SM neutrinos can be constrained by $X$-ray observations, searching for the loop-induced decay of the right-handed neutrinos into active neutrinos and photons.
The allowed parameter space \cite{Adhikari:2016bei,Boyarsky:2018tvu} is such that the mixing with SM neutrinos is tiny, and therefore the keV-neutrinos never reach thermal equilibrium, and thus classical freeze-out does not work.
For heavier sterile neutrinos, the limits from their total decay rate, but also from indirect detection, imply the same situation \cite{Anisimov:2008gg}: Observation implies very small mixing, which means that the sterile neutrinos would never reach thermal equilibrium. 
In this work we shall insist, however, on the production of the DM particle via interactions with the SM particle bath.
We note that this could be achieved if, by any mechanism, the neutrino Yukawa coupling, which is responsible for the strength of its interactions with the SM plasma and its decay, was effectively varying during the early Universe from rather large to very small values. 
To investigate this general idea we consider a type I seesaw extension of the SM embedded in a Froggatt-Nielsen (FN) model \cite{Froggatt:1978nt} and complemented by an additional scalar field $\Sigma$ and global symmetry $U(1)_\Sigma$.

The seesaw extension adds right-handed neutrinos to the theory which couple to the SM via the Yukawa terms.
The FN mechanism introduces an additional $U(1)_\mathrm{FN}$ symmetry under which all or some fermions are charged. 
The symmetry is broken by the vev of the flavon field, which is also charged under the $U(1)_\mathrm{FN}$. 
As a consequence, the Yukawa terms are multiplied by powers of the flavon vev. Usually, the flavon vev is expected to be smaller than the FN scale $\Lambda_\mathrm{FN}$, meaning that the Yukawa terms are effectively suppressed, thus offering an explanation for the fermion flavour structure.
Here, we introduce a new $U(1)_\Sigma$ symmetry under which only the new scalar field $\Sigma$ is charged. At a certain moment, well after $U(1)_\mathrm{FN}$ breaking, the $\Sigma$ scalar gets a vev which breaks the $U(1)_\Sigma$ symmetry. This can cause the global minimum of the scalar potential to shift in field-space, thereby changing the value of the flavon vev.
Therefore, during this phase transition one can think of the Yukawa couplings as effectively varying from one initial value to the value known today. 
If the initial value of the Yukawas was large, then it is possible that the right-handed neutrinos interacted efficiently with the rest of the cosmic plasma, reaching thermal equilibrium and freezing out after the $U(1)_\Sigma$ phase transition occurs. After the phase transition the Yukawa coupling, and thus the mixing, is small enough in order to render the right-handed neutrino stable. 
\\

In the course of this work we will first briefly introduce the seesaw mechanism in \autoref{sec:seesaw}, and in \autoref{sec:model} we describe how the combination of the seesaw and the Froggatt-Nielsen mechanisms can include a dark matter candidate. 
In \autoref{sec:dm-genesis} we go over the interactions and dynamics involved in the thermal equilibrium and freeze-out of seesaw sterile neutrinos and solve the Boltzmann equation to compute the relic abundance. We then conclude in \autoref{sec:conclusio}.

\section{Seesaw mechanism and sterile neutrinos as a dark matter candidate}
\label{sec:seesaw}
Introducing three right-handed Majorana neutrinos $\nu_R$ allows the presence of Yukawa and Majorana terms in the Lagrangian.
The SM Lagrangian is thus extended to include the seesaw Lagrangian,
\begin{align}\label{eq:seesawL_only}
   - {\cal L}_\nu = i \bar{\nu}_R\slashed{\partial}\nu_R + y_\nu \bar{L}\Tilde{\phi}\nu_R + \frac{1}{2} \overline{\nu^c_R} M_R \nu_R  + \mathrm{h.c.},
\end{align}
where $L$ stands for the $SU(2)$ lepton doublets, $\Tilde{\phi}$ is the dual SM Higgs field, $y_\nu$ is the $3\times3$ matrix of neutrino Yukawa couplings and $M_R$ is the $3\times3$ Majorana mass matrix.
In this equation one should think of $L$ and $\nu_R$ as representing all three generations.
After EWSB the vev of the Higgs field $v$ generates the Dirac mass matrix for neutrinos, $m_D = y_\nu \, v/\sqrt{2}$.
Then, in the basis given by $\nu_{\cal M} = (\nu_{L,e},\nu_{L,\mu},\nu_{L,\tau}, \nu_{R,1}^c, \nu_{R,2}^c, \nu_{R,3}^c)^T$ the neutrino mass terms can be written with a $6\times6$ matrix as
\begin{align}
    - {\cal L}_\mathrm{\nu, mass} = \frac{1}{2}\,\overline{\nu^c_{\cal M}}\, {\cal M}_\nu \, \nu_{\cal M} + \mathrm{h.c.} =  \frac{1}{2}\,\overline{\nu^c_{\cal M}}
    \begin{pmatrix}
    0 & m_D \\
    m_D^T & M_R
  \end{pmatrix} \nu_{\cal M} + \mathrm{h.c.},
\end{align}
with the full neutrino mass matrix ${\cal M}_\nu$. 

Notice that the matrices $m_D$ and $M_R$ need not be diagonal.
The central assumption in the seesaw framework is that the Majorana masses in $M_R$ are larger than the Dirac masses in $m_D$.
This is a plausible assumption since $m_D$ is generated at the EW scale by the Higgs mechanism, but $M_R$ likely has its origin at a higher scale of BSM physics, perhaps at the GUT scale.
Then, ${\cal M}_\nu$ can be block diagonalized (to first order approximation) by the unitary $6\times6$ matrix $U$,
\begin{align}\label{eq:active-sterile-theta}
    U = \begin{pmatrix}
    1 & \theta \\
    -\theta^\dagger & 1
  \end{pmatrix}, \qquad \theta = m_D \, M_R^{-1}\,.
\end{align}
The $3\times3$ matrix $\theta$ gives the mixing between left-handed and right-handed neutrinos introduced by the block diagonalization.
The resulting neutrino mass matrix is
\begin{align}
    {\cal M}_\nu' = \begin{pmatrix}
    m' & 0 \\
    0 & M'
  \end{pmatrix}, \qquad m' \approx m_D\,M_R^{-1}\,m_D^T, \qquad M' \approx M_R .
\end{align}
Here the attractiveness of the seesaw framework becomes obvious: if the elements of $M_R$ are much larger than those of $m_D$, then the active neutrino mass eigenvalues $m'$ will be strongly suppressed, thus naturally explaining their tiny size. 
The active and sterile mass matrices can be further diagonalized using appropriate matrices  $V_L$ and $V_R$:
\begin{align}
    m = V_L^T m' V_L = \mathrm{diag}(m_1,m_2,m_3), \qquad M = V_R^T M' V_R = \mathrm{diag}(M_1,M_2,M_3),
\end{align}
and the corresponding light and heavy mass eigenstates are referred to as $\nu = (\nu_1, \nu_2, \nu_3)$ and $N = (N_1, N_2, N_3)$, respectively.

The Yukawa coupling makes sterile neutrinos unavoidably unstable.
To arrive at a rough estimate about the constraints on the Yukawa coupling (and equivalently, the mixing angle) from the longevity condition, we consider one single neutrino generation and we take $m$, $M$ and $y_\nu$ as scalar quantities.
When considering the possible decays it is important to distinguish two different scenarios:
\begin{enumerate}
    \item If the sterile neutrino state is heavier than the real Higgs boson $h$, then the decay $N \rightarrow  h\,\nu$ is kinematically allowed and its width is \cite{Anisimov:2008gg}
    \begin{align}
         \Gamma_{N \rightarrow  h\,\nu} = y_\nu^2\,M/16\pi.
    \end{align}
    If we now demand that the corresponding lifetime be larger than the age of the Universe $t_0 \simeq 10^{41}\,\mathrm{GeV}^{-1}$, we see that $y_\nu$ is constrained by\footnote{Various other decay modes are possible as well, which are important for limits on indirect detection \cite{Anisimov:2008gg,DiBari:2016guw,DeRomeri:2020wng}. The qualitative conclusions do not change if we take those into account.}
    \begin{align}
        y_\nu < 2\cdot 10^{-22}\left( \frac{10\,\mathrm{TeV}}{M} \right)^{1/2}.
    \end{align}
    Clearly, this Yukawa coupling is so tiny that the right-handed neutrinos could never reach thermal equilibrium, thus making production by freeze-out impossible.
    The light mass state generated by this mixing, assuming $M = 10\,\mathrm{TeV}$, is extremely tiny: 
    \begin{align}
        m = \theta^2 M = \left(\frac{y_\nu\,v}{M} \right)^2 M \approx 1.5\,\cdot 10^{-34}\,\mathrm{eV}.
    \end{align}
    
    \item If the sterile neutrino state is lighter than the real Higgs boson, then it  can only decay via its mixing with active neutrinos. In this case, the decay width depends on the decay channels available at a certain mass. For concreteness we take the well known example of a keV scale sterile neutrino. The dominant decay channel is into three active neutrinos and the width is \cite{Pal:1981rm}
    \begin{align}
        \Gamma_{N\rightarrow3\nu} = \frac{G_F^2 M^5}{96\pi^3}\sin^2 \theta  ,
    \end{align}
    with the Fermi constant $G_F$. The longevity condition demands that \cite{Adhikari:2016bei}
    \begin{align}
        \theta^2 < 1.1\cdot 10^{-7} \left( \frac{50\,\mathrm{keV}}{M} \right)^5,
    \end{align}
    which, for $M = 50\,\mathrm{keV}$, implies $y_\nu^2 < 6\cdot 10^{-21}$, thus making production by freeze-out again unfeasible, and production by oscillations an appealing alternative.
    The active neutrino mass in this case is 
    $m = \theta^2 M \approx 5\cdot 10^{-3}\,\mathrm{eV}$.
\end{enumerate}

In this paper, our goal is to show that dark matter sterile neutrinos can be produced in the early Universe by decoupling from thermal equilibrium. This occurs naturally if the sterile neutrinos have a sizeble Yukawa coupling at the time of DM generation, but becomes small afterwards, thus keeping the sterile neutrinos stable on cosmological time scales.

\section{Seesaw-Froggatt-Nielsen models for sterile neutrino dark matter}
\label{sec:model}

The Froggatt-Nielsen mechanism was introduced to explain the flavour asymmetry in the masses of fermions.
It postulates the existence of a new $U(1)_\mathrm{FN}$ symmetry and a scalar field, called the flavon $\Theta$, which encodes a heavy hidden sector that is integrated out below a certain scale $\Lfn$. 
The left- and right-handed fermions of the $i$'th generation have the FN charges $f_i$ and $g_i$, respectively, and $\Theta$ has the FN charge $-1$.
For the theory to be invariant under $U(1)_\mathrm{FN}$, the Yukawa terms have to be modified in the following way:
\begin{align}\label{eq:FN-yukawa-term}
    y_{ij}\, \overline{\psi_L}_i\,\phi \, \psi_{R\,j} \longrightarrow  y_{ij}\,\overline{\psi_L}_i\,\phi \,\psi_{R\,j}\, \left(\frac{\Theta}{\Lfn}\right)^{f_i + g_j}.
\end{align}
Then, at some moment the $U(1)_\mathrm{FN}$ symmetry is broken by the vev of the flavon, for which we define
\begin{align}\label{eq:lambda}
    \lambda := \frac{\langle \Theta \rangle}{\Lfn}.
\end{align}
This means that, if $\lambda<1$, the Yukawa terms in Eq.\ \eqref{eq:FN-yukawa-term} will be suppressed by $\lambda$ to the power of the sum of the FN charges of the fermions involved. 
The hierarchy in the masses is therefore explained by the breaking of the $U(1)_\mathrm{FN}$ symmetry and not by a strong hierarchy in the Yukawa couplings $y_{ij}$, which, in this framework, may all be of order unity.
This mechanism is known to work well for the quark sector, where $\lambda = 0.22$ has been related to the Cabibbo angle of the CKM matrix. 
If it is a correct description for the origin of the quark flavour structure, and right-handed neutrinos exist, then it would be natural to expect that the FN mechanism also applies to leptons and, by inclusion, to neutrinos.

In this work we want to claim that, within a setting of dynamical Yukawa couplings, RH neutrinos involved in the type I seesaw mechanism may constitute the dark matter density and be produced by freeze-out from the thermal bath.
To investigate this idea, we propose the following toy-model realization: consider a type I seesaw extension to the SM embedded in a FN model and add to it an additional global symmetry (e.g. $U(1)$ or $Z_2$) under which a new scalar field $\Sigma$ is charged.
As we will see, the key concept is the dynamical suppression of the Yukawa coupling and Majorana mass, which is realized by the breaking of the new global symmetry induced the vev of the $\Sigma$ field.
The SM is extended by the following elements:
\begin{itemize}
	\item three right-handed Majorana neutrinos, $N_{1,2,3}$,
	\item a scalar field, the FN flavon $\Theta$, and a global $U(1)_\mathrm{FN}$ flavour symmetry (the UV complete theory contains the FN messengers).
	\item a scalar field $\Sigma$, and a global $U(1)_\Sigma$ symmetry
\end{itemize}
Apart from kinetic terms and the scalar potential, which we will discuss in detail shortly, the only additional terms for the Lagrangian are the Majorana mass term and the neutrino Yukawa term, both of which include the flavon.
One of the sterile neutrinos will play the role of the dark matter particle (denoted simply by $N$ from now on) while the other two could be responsible for generating the masses of active SM neutrinos and could also be involved in leptogenesis.
From this point forward we will consider only one lepton generation, as this is all that is relevant for DM production. 
This is the case because, as we will see, DM production will be driven solely by the Yukawa term and we can always switch to a basis in which the Yukawa matrix is diagonal.
In this basis the DM neutrino $N$ couples only to one lepton doublet, which we take as the first generation doublet. 
We assign the $U(1)_\mathrm{FN}$ charges as in \autoref{tab:FNcharges}.

\begin{table}[tbp]
\begin{center}
\begin{tabular}{lccccc}
\toprule
    Field      & $L$  & $e_R$ & $N$ & $\Theta$ \\
\midrule  
    $U(1)_\mathrm{FN}$ Charge &  $q_L$   &  $q_R$   &  $q_N$ & $-1$ \\
\bottomrule
\end{tabular}
\caption{\label{tab:FNcharges} FN charges in our simplified model.}
\end{center}
\end{table} 

\begin{figure}[t]
\begin{center}
	\includegraphics[width=0.9\textwidth]{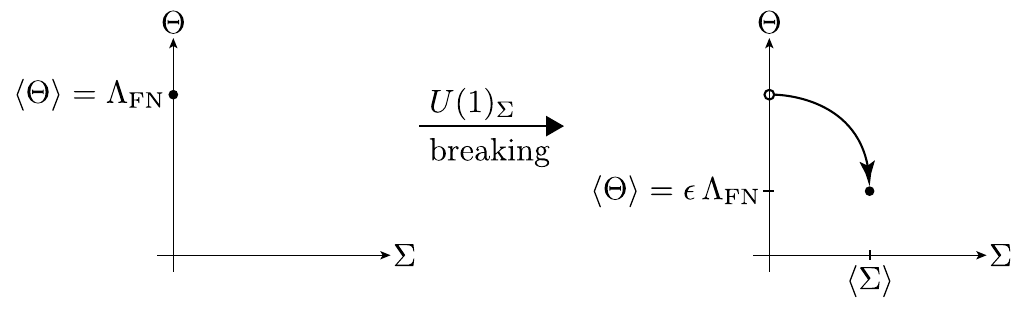}
	\caption{Location of the minimum of the scalar potential in field space and shift in the value of the flavon vev caused by the $U(1)_\Sigma$ symmetry breaking.}
	\label{fig:sigma-pt}
\end{center}
\end{figure}

The relevant part of the $U(1)_\mathrm{FN}$ symmetric Lagrangian is
\begin{align}\label{eq:seesawL}
  y_e\left( \frac{\Theta}{\Lfn} \right)^{q_L+q_R} \bar{L}\,{\phi}\,e_R \,+\, y_\nu \left( \frac{\Theta}{\Lfn} \right)^{q_L+q_N} \bar{L}\,\Tilde{\phi}\,N \,+\, \frac{1}{2} M_R \left( \frac{\Theta}{\Lfn} \right)^{2q_N} \overline{N^c} N  .
\end{align}
The main idea of the FN mechanism is that all of the Yukawa couplings are (close to) order unity.
Once the $U(1)_\mathrm{FN}$ symmetry is broken by the vev of the flavon $\langle \Theta \rangle$, and if $\lambda<1$ (see Eq. \eqref{eq:lambda}), the Yukawa interactions are suppressed by powers of $\lambda$.
However, another possibility is that $\langle \Theta \rangle \approx \Lambda_\mathrm{FN}$, i.e.\ $\lambda\approx1$, which would mean that the Yukawa couplings are actually not suppressed and may be of order 1.
In particular, the sterile neutrino Yukawa interactions would be strong enough to keep them in thermal equilibrium with the cosmic plasma.
Whether the sterile neutrinos were already present in large numbers in the plasma or not, is irrelevant.
If they were present, e.g.\ after being produced during  reheating, then they were kept in equilibrium by their unsuppressed Yukawa coupling.
If they were not present initially, then they were produced in the plasma by their Yukawa interactions and thermalized quickly.
At this stage, the minimum of the scalar potential $V(\phi,\,\Theta,\,\Sigma)$ in field space is located at $(\phi,\,\Theta,\,\Sigma) = (0,\,\langle \Theta \rangle,\,0)$, with $\langle \Theta \rangle \approx \Lambda_\mathrm{FN}$.
This is where the new scalar field $\Sigma$ comes into play. It is the only field charged under the $U(1)_\Sigma$ symmetry. It too could acquire a vev and break the $U(1)_\Sigma$ symmetry. After such a phase transition, the location of the minimum of the potential is shifted in field space,
\begin{align}
	(\langle \phi \rangle,\,\langle \Theta \rangle,\,\langle \Sigma \rangle) = (0,\, \Lambda_\mathrm{FN},\,0) \stackrel{U(1)_\Sigma}{\longrightarrow} (\langle \phi \rangle,\,\langle \Theta \rangle,\,\langle \Sigma \rangle) = (0,\,v_\theta,\,v_\sigma),
\end{align}
where the new value of the flavon vev $v_\theta$ is different than before the $U(1)_\Sigma$ phase transition.
This is the key point: if the $v_\theta = \epsilon\,\Lambda_\mathrm{FN}$ with $\epsilon<1$, then after the phase transition the Yukawa couplings become suppressed by powers of $\epsilon$.
In terms of \cref{eq:lambda}, the $\lambda$-parameter can be understood as describing a path in field space with boundaries given by,
\begin{align}
	\lambda(\langle \Sigma \rangle) =
	\begin{cases}
	1,\quad & \mathrm{for} \quad \langle \Sigma \rangle = 0 \\
	\epsilon,\quad & \mathrm{for} \quad \langle \Sigma \rangle = \vs.
	\end{cases}
\end{align}
The precise trajectory in field space during the phase transition is not really important.
What matters is that the flavon vev has different values in the different phases, as sketched in \cref{fig:sigma-pt}.
This is very similar to the situation described by Baldes, Konstandin and Servant in Section 4.1 of \cite{Baldes:2016gaf}; however, they consider the quark sector in the context of EW baryogenesis.
Depending on the FN charges, the suppression could be enough to freeze the DM neutrino interactions out, leaving behind a DM neutrino relic abundance similar to the usual WIMP paradigm.
Notice that just like the Yukawa couplings, this mechanism will also suppress the Majorana mass of the sterile neutrinos. Effects of phase transitions in the production of sterile neutrino dark matter have been discussed previously, see e.g.\ \cite{Bezrukov:2017ike,Hansen:2017rxr,DiBari:2020bvn,DeRomeri:2020wng,Heurtier:2019beu,Baker:2017zwx}. Our focus on freeze-out production of a heavy sterile neutrino differs from those previous works.
Whether the phase transition can indeed shift the value of the flavon vev depends on the parameters of the scalar potential, which we discuss hereinafter, following closely the treatment in \cite{Baldes:2016gaf}.
\begin{center}
	\textbf{The scalar potential}
\end{center} 

We assume that the scales $\Lfn$ and $\Lsi$ are much larger than $\Lew$ and the couplings of the $\Theta$ and the $\Sigma$ fields to the SM Higgs boson are small, such that the dynamics of the $\Theta$ and the $\Sigma$ fields can be considered as decoupled from the Higgs boson. 
Then, the relevant part of scalar potential reads
\begin{align}
	V(\Theta,\,\Sigma) = \mut\,\Theta^\dagger\Theta + \lt\,(\Theta^\dagger\Theta)^2 + \mus\,\Sigma^\dagger\Sigma + \ls\,(\Sigma^\dagger\Sigma)^2 + \lts\,(\Theta^\dagger\Theta)(\Sigma^\dagger\Sigma).
\end{align}
Without loss of generality, we assume that both fields acquire their vev's only along their real components and substitute
\begin{align}
	\Theta \longrightarrow \frac{1}{\sqrt{2}}\,\theta, \qquad \qquad \Sigma \longrightarrow \frac{1}{\sqrt{2}}\,\sigma,
\end{align}
leading to 
\begin{align}
	V(\theta, \sigma) = \frac{\mut}{2}\,\theta^2 + \frac{\lt}{4}\,\theta^4 + \frac{\mus}{2}\,\sigma^2 + \frac{\ls}{4}\,\sigma^4 + \frac{\lts}{4}\,\theta^2\sigma^2.
\end{align}
The minimum of the potential is given by the zeros of the field space gradient:
\begin{align}
	\nabla_{\theta,\sigma} V =
	\begin{pmatrix}
	\theta\,(\mut + \lt\,\theta^2 + \frac{\lts}{2}\,\sigma^2) \\
	\sigma\,(\mus + \ls\,\sigma^2 + \frac{\lts}{2}\,\theta^2) 
	\end{pmatrix}\Bigg\rvert_{(\vt,\vs)} \stackrel{!}{=} 
	\begin{pmatrix}
		0\\
		0
	\end{pmatrix}.
\end{align}
We recognise the following cases:
\begin{enumerate}[(i)]
	\item \textbf{Before the $U(1)_\Sigma$ phase transition:} At temperatures $\Lambda_\Sigma<T<\Lfn$ we demand
		\begin{align}
			\vt=\Lfn, \qquad \text{and} \qquad \vs = 0.
		\end{align}
		Thus, the first component of $\nabla_{\theta,\sigma}V = 0$ implies
		\begin{align} \label{eq:mut}
			\mut = - \lt \, \Lfn^2.
		\end{align}
	\item \textbf{After $U(1)_\Sigma$ symmetry breaking:}\\
		At temperatures $\Lew < T < \Lambda_\Sigma$ we demand
		\begin{align}
			\vt=\epsilon \, \Lfn, \qquad \text{and} \qquad \vs \neq 0,
		\end{align}
		with $0<\epsilon<1$. Thus, the first component of $\nabla_{\theta,\sigma}V = 0$ implies
		\begin{align}
			\mut + \lt \, \epsilon^2\,\Lfn^2 + \frac{\lts}{2}\, \vs^2 = 0. 
		\end{align}	
		With Eq. \eqref{eq:mut} this is equivalent to 
		\begin{align}\label{eq:scalar-scales}
			\underbrace{\left(1-\epsilon^2 \right)}_{\sim 1} \left(\frac{\Lfn}{\vs}\right)^2 = \frac{\lts}{2\, \lt} \qquad 
			\Rightarrow	\qquad
			\left(\frac{\Lfn}{\vs}\right)^2 \approx \frac{\lts}{2\, \lt}.
		\end{align}
\end{enumerate}
Next we compute the mass parameters:
\begin{align}
		m_{\theta \theta}^2 &= \partial_\theta^2V = \mut + 3\lt\vt^2+\frac{1}{2}\lts \vs^2,\\
		m_{\sigma \sigma}^2 &= \partial_\sigma^2V = \mus + 3\ls\vs^2+\frac{1}{2}\lts \vt^2, \\
		m_{\theta \sigma}^2 &= \partial_\theta\partial_\sigma V = \lts \vs \vt.
\end{align}
The mass matrix is
\begin{align}
	M = \begin{pmatrix}
		m_{\theta \theta}^2 & m_{\theta \sigma}^2 \\
		m_{\theta \sigma}^2 & m_{\sigma \sigma}^2
	\end{pmatrix} 
\end{align}
and its corresponding mass eigenvalues are
\begin{align}
	m_\sigma^2 = \ls \vs^2 + \lt \vt^2 + \sqrt{(\ls \vs^2 - \lt \vt^2)^2 + (\lts \vt \vs)^2}, \\
	m_\theta^2 = \ls \vs^2 + \lt \vt^2 - \sqrt{(\ls \vs^2 - \lt \vt^2)^2 + (\lts \vt \vs)^2}. 
\end{align}
The mixing angle between $\theta$ and $\sigma$ is given by 
\begin{align}
		\tan(2\alpha) = \frac{\lts \vt \vs}{\ls \vs^2 - \lt \vt^2} = \frac{\lts \vs \, \epsilon \, \Lfn}{\ls \vs^2 - \lt \, \epsilon^2\, \Lfn^2}.
\end{align}
As we will see in a moment, in our scenario, the mixing angle is automatically very small, implying 
\begin{align}
	\lts \vt \vs \ll |\ls \vs^2 - \lt \vt^2|,
\end{align}
which means that the mass eigenvalues can be approximated by
\begin{align}
	m_\sigma^2 = 2\,\ls \vs^2, \\
	m_\theta^2 = 2\,\lt \vt^2. \label{eq:flavon-mass}
\end{align}
Now, for the stationary point $(\theta,\, \sigma) = (\vt,\,\vs) = (\epsilon\,\Lfn,\, \vs )$ to be indeed a minimum of the potential, the determinant of the mass matrix must be positive definite:
\begin{align}
	\det(M) = \det\begin{pmatrix}
		\partial_\theta^2V & \partial_\sigma\partial_\theta V \\
		\partial_\sigma\partial_\theta V & \partial_\sigma^2V
	\end{pmatrix} \Bigg\rvert_{(\vt,\vs)} >0,
\end{align}
which results in
\begin{align}
	4 \, \ls \lt > \lts^2.
\end{align}
From this relation and with Eq. \eqref{eq:scalar-scales}, we derive constraints on the coupling constants
\begin{align}
	\lts < 2 \, \ls \left( \frac{\vs}{\Lfn}\right )^2, \label{eq:lts-bound} \\
	\lt < \ls \left( \frac{\vs}{\Lfn}\right )^4. \label{eq:lt-bound}
\end{align}
These constraints consistently imply the smallness of the mixing angle.\ 
Finally, we obtain a bound on the mass of the flavon by combining Eqs. \eqref{eq:flavon-mass} and \eqref{eq:lt-bound},
\begin{align} \label{eq:flavon-mass-bound}
	m_\theta^2  < 2  \ls \,\vs^2\,\epsilon^2 \, \left(\frac{\vs}{\Lfn}\right)^2. 
\end{align}
From this analysis we have learned that, if \cref{eq:lts-bound,eq:lt-bound,eq:flavon-mass-bound} hold, then the $U(1)_\Sigma$ phase transition can cause a shift in the value of the flavon vev.
It is crucial that, for the effective field theory approach to be applicable to the Yukawa terms, the $\uone$ symmetry breaks before (i.e.\,at a higher temperature than) the $U(1)_\Sigma$ symmetry in cosmological history.
Since, as a rough estimate, we can relate $\vs \sim \Lambda_\Sigma$, the ratio $(\vs/\Lfn)$, which appears in all the afore mentioned bounds, should be smaller than 1, i.e.
\begin{align}
	\delta := \left(\frac{\vs}{\Lfn}\right) \sim \left(\frac{\Lambda_\Sigma}{\Lfn}\right)<1.
\end{align}
For a specific FN model, the mass of the flavon can be constrained by experiments looking for flavour violating transitions in the lepton sector, e.g.\,exotic $\mu$ decays.
From such measurements and \cref{eq:flavon-mass-bound}, one can obtain a lower bound for $\Lambda_\Sigma$, and, since the critical temperature $T_c$ of the phase transition is expected to be close to this scale, this bound also approximately applies for $T_c$ as well,
\begin{align}\label{eq:Tc-bound}
	m_\theta\,\epsilon^{-1}\,\delta^{-1} < \Lambda_\Sigma \sim T_c.
\end{align} 

\begin{center}
	\textbf{Dark Matter toy model}
\end{center}
From the discussion of the scalar potential, we have seen that within this framework the $U(1)_\Sigma$ phase transition is capable of shifting the value of the flavon vev as $\langle \Theta \rangle : \Lfn \rightarrow \epsilon\,\Lfn$.
As a consequence, the neutrino Yukawa coupling (and Majorana mass) may be considered as effectively varying during the phase transition,
\begin{align}\label{eq:yeff}
	y_\mathrm{eff} = y_\nu \, \left( \frac{\langle \Theta \rangle}{\Lfn} \right)^{q_L+q_N} =
		\begin{cases}
			y_i = y_\nu, &\mathrm{for}\quad T>T_c \\
			y_f = y_\nu\,\epsilon^{\,q_L+q_N}, &\mathrm{for}\quad T<T_c 
		\end{cases},
\end{align}

\begin{align}\label{eq:initial_final_M}
	M_\mathrm{eff} = M_R \, \left( \frac{\langle \Theta \rangle}{\Lfn} \right)^{2q_N} =
		\begin{cases}
			M_i = M_R, &\mathrm{for}\quad  T>T_c \\
			M_f = M_R \,\epsilon^{\,2q_N}, &\mathrm{for}\quad T<T_c 
		\end{cases}.
\end{align}

Here we refer to the values of the Yukawa coupling and DM neutrino Majorana mass before and after the $U(1)_\Sigma$ phase transition with the indices $i$ and $f$, for \textit{initial} and \textit{final} respectively.
The temperature $T_c$ is the temperature at which the $U(1)_\Sigma$ phase transition occurs.

In a full FN model, we would compute the flavon-lepton couplings and with experimental data derive a constrain for the flavon mass, which then results in a lower bound for the phase transition temperature $T_c$ from \cref{eq:Tc-bound}.
However, that would go far beyond the scope of this work, as we are here only concerned with a proof of principle for the freeze-out mechanism as a means of production for sterile neutrino DM.
Instead, in this work we will simply assume a flavon mass in the GeV range (the flavon-lepton couplings could certainly be very small), and leave the concrete realization in a full FN model for future work.
Furthermore, we set for concreteness $\delta = 0.1$ and $\epsilon = 0.01$.
As we will later see, successful DM production requires that $M_i \sim 35\,T_c$, which means that the lower bound from \cref{eq:Tc-bound} on $T_c$ also places a lower bound on the unsuppressed Majorana mass of DM neutrinos, namely $M_i>10\,\mathrm{TeV}$.

Since the DM neutrinos are heavier than the Higgs, the tree level decay is allowed.
To comply with the longevity condition we must demand that the DM neutrino be stable on cosmological time scales. 
This results in a condition on the FN charges. 
Other conditions arise from the other two terms in Eq.\ \eqref{eq:seesawL}:
\begin{enumerate}
	\item \textbf{The electron mass:} With $y_e$ of order unity, the unsuppressed electron mass would be $y_e\,v/\sqrt{2} \sim 10^{2}\,\mathrm{GeV}$. This has to be suppressed to the level of $m_e\approx 5\cdot 10^{-4}\,\mathrm{GeV}$,
		\begin{align}\label{eq:FNcond1}
			\epsilon^{\,q_L+q_R}\, 10^{2}\,\mathrm{GeV} \approx 5\cdot 10^{-4}\,\mathrm{GeV} \qquad \Rightarrow \qquad q_L+q_R \approx \frac{-5}{\log_{10}(\epsilon)}.
		\end{align}
	\item \textbf{The Majorana mass:} As explained above, the suppressed Majorana mass is bounded from below by $M_i>10\,\mathrm{TeV}$ and is UV-unconstrained.
		The final DM sterile neutrino is heavier than the Higgs,
		\begin{align}\label{eq:FNcond2}
			M_f = \epsilon^{\,2q_N}\, M_i \gtrsim 2\cdot10^{2}\,\mathrm{GeV} \qquad \Rightarrow \qquad q_N\lesssim \frac{\log_{10}(2\cdot10^2)-\log_{10}\left(\frac{M_i}{\mathrm{GeV}}\right)}{2\log_{10}(\epsilon)}.
		\end{align}
	\item \textbf{The longevity of DM:} DM must be stable on cosmological time scales, i.e.\  $\tau_\mathrm{DM}=\Gamma_{N\rightarrow h\nu}^{-1}>t_0\approx7\cdot10^{41}\,\mathrm{GeV}^{-1}$.
	With $\Gamma_{N\rightarrow h\nu} = y_f^2 M_f/16\pi$ we get 
		\begin{align}\label{eq:FNcond3}
			\frac{(y_i \,\epsilon^{\,q_L+q_N})^2M_i\,\epsilon^{\,2q_N}}{16\pi} < \frac{1}{7}\cdot10^{-41}\,\mathrm{GeV}
			\quad \Rightarrow \quad
			q_L+2q_N > \frac{\log_{10}\left( \frac{16\pi}{7}\cdot10^{-41} \left(\frac{1}{y_i} \right)^2 \left(\frac{\mathrm{GeV}}{M_i}\right)\right)}{2\,\log_{10}(\epsilon)}.
		\end{align}
\end{enumerate}
In order to satisfy the conditions, the FN charge of $e_R$ must be negative.
This can be seen as a positive side-effect of the conditions above, because it means that the electron Yukawa term will be multiplied by a smaller power of $\Theta$ fields.

We can find appealing choices of charges for our fields by solving the problem 
\begin{align}\label{eq:optimazation}
	\underset{q_L,q_R,q_N \, \in \, \mathbb{Z} }{\text{minimize}} \left(|q_L|+|q_R|+|q_N|\right)
\end{align}
under the conditions from \cref{eq:FNcond1,eq:FNcond2,eq:FNcond3}.
For the concrete models discussed in this work, we fix the parameters $y_i$ and $\epsilon$ to 
\begin{align}
	y_i = 0.1, 	\qquad	\epsilon = 0.01.
\end{align}
Solving the optimization problem \cref{eq:optimazation} for different Majorana mass ranges results in the three different models given in \cref{tab:FN-configurations}. 
We find that, for our choice of $y_i$ and $\epsilon$, the conditions \cref{eq:FNcond1,eq:FNcond3} imply that $q_L+q_R=2$ and $q_L+q_N=11$, while $q_N$ is directly determined by \cref{eq:FNcond2}. 
Each model is defined by its FN charges. In each model we allow a range of initial masses $M_i$, which translates in a range of $M_f$ values. This range is the same for all three classes of models, namely $M_f$ starts at $10^4\,$GeV and goes up to $10^8\,$GeV. 
\begin{table}[t]
\begin{center}
\begin{tabular}{ccccc}
\toprule
Model &  M1  &  M2   &  M3  \\
\toprule
  $M_i/\,$GeV & $\left[10^{4},10^{8}\right)$  & $\left[10^{8},10^{12}\right)$ & $\left[10^{12},10^{16}\right)$ \\
\midrule  
    $q_L$ &  $11$  &  $10$   &  $9$  \\
\midrule
	$q_R$ & $-9$   & $-8$   & $-7$  \\
\midrule
	$q_N$ &  $0$   &  $1$   &  $2$  \\
\bottomrule
\end{tabular}
\caption{We define three classes of FN models for different $M_i$ ranges. The ranges of $M_i$  include the lower bound but exclude the upper bound, as indicated by the square and round brackets. The corresponding range for $M_f$ is $[10^4\,\mathrm{GeV},10^8\,\mathrm{GeV})$ for all three classes of models.
The configurations of the FN charges satisfy the conditions \cref{eq:FNcond1,eq:FNcond2,eq:FNcond3} for our choice of $\epsilon =0.01$ and $y_\nu =0.1$. For smaller values of $\epsilon$ the charges can take even smaller numerical values.}
\label{tab:FN-configurations}
\end{center}
\end{table}
Our choice of $\epsilon = 0.01$ may mean that our flavon is possibly not the same as the CKM flavon, whose vev is typically related to the Cabbibo angle, or $\epsilon \approx 0.22$.
This would not be a problem, since the flavour breaking structure in leptonic sector does not need to be identical to that in the quark sector.

\section{Dark matter genesis}
\label{sec:dm-genesis}

Our goal is to determine the relic abundance of dark matter neutrinos produced for the different models described in the previous section and given in \cref{tab:FN-configurations}.
To this end, we solve the Boltzmann equation for the number density of DM neutrinos in the early Universe.

As mentioned in \cref{sec:model}, we assume that at high temperatures, in the $U(1)_\Sigma$ symmetric phase, the vev of the flavon field is $\langle \Theta \rangle \approx \Lfn$, implying that before the symmetry breaking phase transition the Yukawa coupling and the Majorana mass are large and unsuppressed.
The sterile neutrinos are kept in thermal equilibrium by the decays and inverse decays allowed by the Yukawa coupling.
Other interactions, such as $2 \leftrightarrow 2$ scatterings involving gauge bosons and quarks are also allowed at tree-level.
However, these are only relevant in the relativistic regime.
As soon as the phase transition occurs, the vev of the flavon is shifted as $\langle \Theta \rangle : \Lfn \rightarrow \epsilon\,\Lfn$ and the interaction rate $\gamma$, which is proportional to the square of the Yukawa coupling, gets strongly suppressed.
The condition under which a particle species with an interaction rate $\gamma$ with the plasma  and an equilibrium number density $n_\mathrm{eq}$ is in thermal equilibrium, is given by
\begin{align}
	\frac{\gamma/n_\mathrm{eq}}{H} 
	\begin{cases}
		> 1, \qquad \text{in thermal equilibrium},\\
		< 1, \qquad \text{out of thermal equilibrium},
	\end{cases}
\end{align}
where $H$ stands for the Hubble rate.
The sudden change in $\gamma$ due to the $U(1)_\Sigma$ phase transition causes the condition to be violated at the time of the transition and almost immediately induces the freeze-out of the DM neutrinos. 
We are not interested in the specific dynamics of the phase transition at this point.
What matters to us are the different states before and after the phase transition, as described by \cref{eq:yeff}.
For our computations, however, we need a specific parametrization for both $y_{\mathrm{eff}}$ and $M_{\mathrm{eff}}$.
A simple and generic parametrization which encodes the relevant behaviour 
(essentially a slightly smoothed step function) for us is given by 
\begin{eqnarray}
    y_\mathrm{eff}(z) &=& \frac{1}{2}\,\left[(y_i-y_f)\, \tanh\left(\left(1-\frac{z}{z_c}\right)\frac{1}{\tau}\right)+y_i+y_f \right], \\
    M_{R,\mathrm{eff}}(z) &=& \frac{1}{2}\,\left[(M_i-M_f)\, \tanh\left(\left(1-\frac{z}{z_c}\right)\frac{1}{\tau}\right)+M_i+M_f \right],
\end{eqnarray} 
where the indices $i$ and $f$ stand for initial and final values, i.e.\  before and after the phase transition. A measure for the time the phase transition takes is $\tau$ (which in our computations we set to $\tau = 0.001$), and $T_c$ is its  critical temperature, with $z_c = M_i/T_c$. 

\begin{figure}[t!]
\centering
\includegraphics[width=\textwidth]{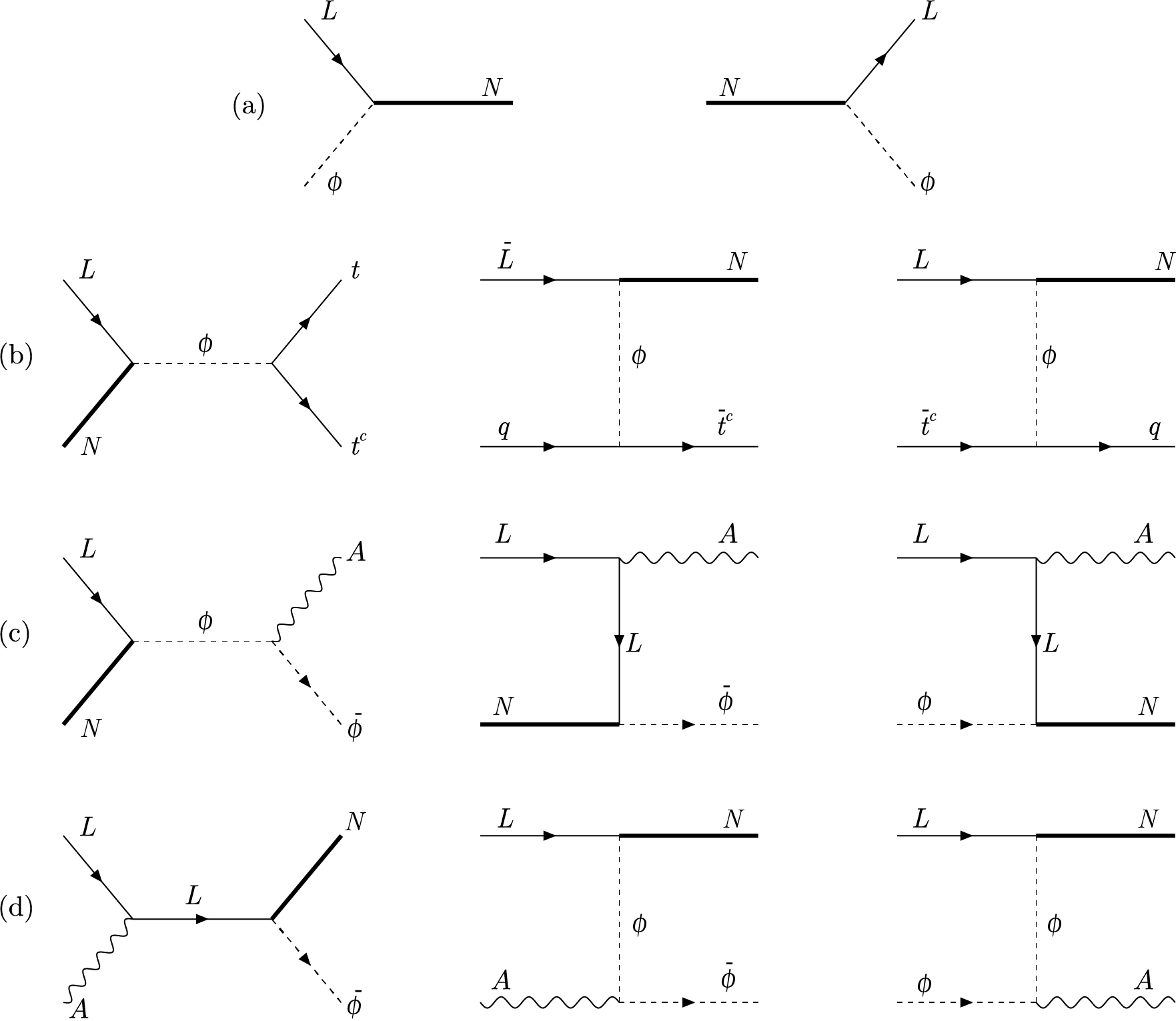}
\caption{\label{fig:decay-and-scatt} Tree-level interactions allowed to sterile neutrinos $N$ in a seesaw framework. Gauge bosons are denoted as $A$.}
\end{figure}

Since our computations take place in the very early Universe, before the electroweak phase transition, the particles involved in the DM interactions will only have thermal masses.
It is important to take these into account because at high temperatures the thermal masses can have sizeable values and in particular, not including them would lead to overestimating the contribution from $2 \leftrightarrow 2$ scattering processes compared to  decays \cite{Giudice2004}. 
The interactions that are relevant for thermal equilibrium and freeze-out of DM sterile neutrinos are those that are allowed by the Yukawa coupling and change the number of DM neutrinos.
These are the decays and inverse decays of $N$ (shown in \cref{fig:decay-and-scatt} (a)), scatterings involving quarks (shown in \cref{fig:decay-and-scatt} (b)) and scatterings involving bosons (shown in \cref{fig:decay-and-scatt} (c, d)).
Note that  $2 \leftrightarrow 2$ scatterings where the sterile neutrinos appear as virtual particles in the propagator are not relevant to us, because they do not change the number of sterile neutrinos\footnote{In contrast, these processes are very much relevant in the context of thermal leptogenesis.}.

\begin{figure}[t!]
\begin{center}
  \includegraphics[width=1\textwidth]{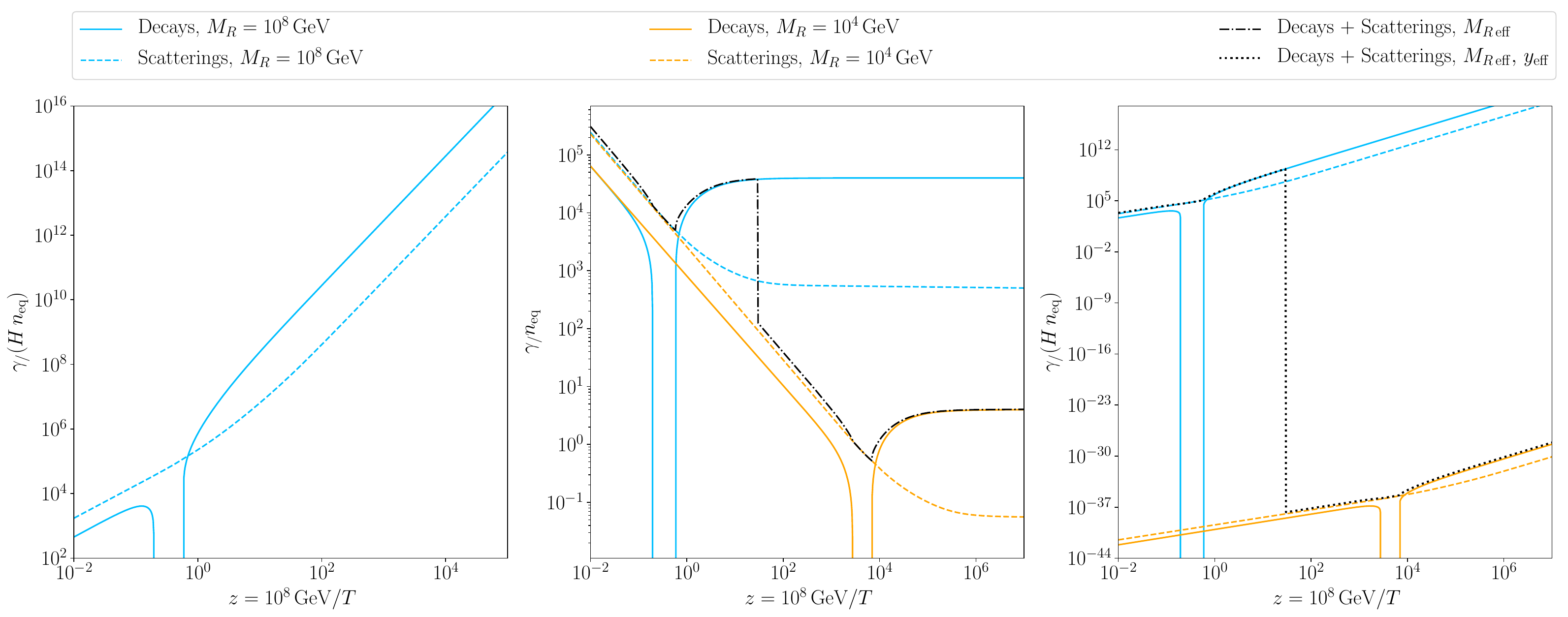}
  \caption{\textit{Left:} The decay and scattering rate for sterile neutrinos with parameters $M_i = 10^{8}\,$GeV and $y_i = 0.1$, without the effects of a coupling to a flavon and the $U(1)_\Sigma$ phase transition. For other values of $M_i$ the rates maintain the same shape but are shifted along the vertical axis (if one adjusts the horizontal axis accordingly to $z=M_i/T$). It is easy to see that the decays dominate in the non-relativistic regime while scatterings are more relevant for $z<1$. \textit{Center:} The decay and scattering rates for two different fixed masses shown in solid and dashed lines, respectively. The smaller mass is chosen such that it coincides with the suppressed Majorana mass $M_f$ for the larger initial mass $M_i = 10^{8}\,$GeV (see \cref{tab:FN-configurations} and \cref{eq:initial_final_M}). The dash-dotted line shows the jump in the total rate due to the suppression of the Majorana mass (leaving out the suppression of the Yukawa coupling) caused by the $U(1)_\Sigma$ phase transition. Before and after the phase transition the effective total rate coincides with the corresponding total rate for different fixed masses. \textit{Right:} Same as center panel, but also showing the effect of the phase transition on the Yukawa coupling, which enters $\gamma$ quadratically. Here, the rates are also divided by the Hubble rate $H$, just as in \cref{eq:boltzi1}.}
  \label{fig:suppressed_decay_rate}
\end{center}
\end{figure}

The Boltzmann equation for the production of DM in this scenario, formulated for the yield $Y=n/s$ (with the entropy density $s$) is given by ($z = M_i/T$) 
\begin{align}\label{eq:boltzi1}
	\frac{dY}{dz}=\frac{1}{z}\left( \frac{\langle\gamma\rangle}{H} \right)(Y_\mathrm{eq} - Y)\,,
\end{align}
where $\langle\gamma\rangle = \gamma/n_\mathrm{eq}$ and $\gamma$ receives contributions from decays and scatterings, i.e.\ $\gamma = \gamma_\mathrm{decay} + \gamma_\mathrm{scatt}$.
For both $\gamma_\mathrm{decay} $ and $ \gamma_\mathrm{scatt}$ we use the cross sections and formulae given in Ref.\ \cite{Giudice2004}, which include thermal corrections and  running of the SM couplings.
Both interaction rates, normalized by $H\,n_\mathrm{eq}$, are shown in the left panel of \cref{fig:suppressed_decay_rate}.
Clearly, decays dominate in the non-relativistic regime, i.e.\ for $z>1$, while the sum of all $2\rightarrow 2$ scatterings delivers a larger contribution at $z<1$. 
Notice that $\langle \gamma\rangle/H$  grows with decreasing temperature, which means that, without the assistance of the $U(1)_\Sigma$ phase transition, instead of departing from equilibrium, the sterile neutrinos would interact ever more strongly with the cosmic plasma. 
 The increase in $\langle \gamma\rangle/H$ with $z$ does not come from $\langle \gamma\rangle$, which stays constant with respect to $z$ for $z>1$, but from $H$, which is proportional to $T^2$ and therefore decreases with $z$. 
The small gap where the decay rate vanishes occurs in the range of temperatures for which $m_\phi(T)-m_L(T) < M_{R,\mathrm{eff}} < m_\phi(T)+m_L(T)$, where $m_\phi(T)$ and $m_L(T)$ are the thermal masses of $\phi$ and $L$, respectively \cite{Giudice2004}. In this temperature range no two-body decays involving $N$ are allowed.

In our framework, the $U(1)_\Sigma$ phase transition causes a sudden suppression of the Majorana mass and the Yukawa coupling.
The impact of the $U(1)_\Sigma$ phase transition on $\gamma_\mathrm{decay}$ and $\gamma_\mathrm{scatt}$ is shown in the center and right panels of \cref{fig:suppressed_decay_rate}.
To understand the effects of the phase transition on $\gamma$ it is useful to first analyse the change in the Majorana mass only.
The center panel on \cref{fig:suppressed_decay_rate} shows in solid and dashed lines the decay and scattering rates for two constant Majorana masses.
Notice that the horizontal $z$-scale refers to the larger mass (in this case, $M_i = 10^{8}\,$GeV) and the rates are not divided by the Hubble parameter $H$ (in contrast to the left and right panels) in order to make the characteristics of the curves easier to distinguish.
The dash-dotted line in this panel is the effective total DM neutrino interaction rate for the model M2 from \cref{tab:FN-configurations} with $M_i = 10^8\,$GeV.
One clearly sees that, prior to the $U(1)_\Sigma$ phase transition, the effective total rate follows that for $M_i = 10^8\,$GeV.
The phase transition occurs for this example at $z_c = 30$ and causes the mass to get suppressed as given by \cref{eq:initial_final_M}, i.e.\ to the value of $M_f = 10^4\,$GeV.
From $z_c$ onward, the effective total rate just follows the total rate for the fixed final mass $M_f$.
In this example, the phase transition occurs when the DM neutrinos are non-relativistic, but immediately after the phase transition they effectively become much lighter while the temperature of the Universe is still $T\sim 10^7\,$GeV.
This means that the DM neutrinos are relativistic again until $T\sim M_f$.
For pedagogical purposes, the center panel of \cref{fig:suppressed_decay_rate} disregards the suppression of the Yukawa coupling.
This effect is accounted for in the right panel, where the rates are also divided by $H$ to also account for the expansion of the Universe.
Here it becomes clear that the effect of the suppression of the Yukawa coupling is much more dramatic than that of the Majorana mass.
Indeed, looking at the gap in orders of magnitude before and after the $U(1)_\Sigma$ phase transition, it is obvious that the induced decoupling from equilibrium, i.e.\ freeze-out, is inevitable.
It is this drastic suppression that guarantees the longevity of the DM. 

A peculiarity occurs with the equilibrium yield $Y_\mathrm{eq}$, which implicitly depends on the mass of the DM particle,
\begin{align}
	Y_\mathrm{eq} \sim z^2\,\mathrm{K}_2(z), \qquad \text{with} \qquad z=M_{R,\mathrm{eff}}/T\,.
\end{align}
Here $\mathrm{K}_2(z)$ is the modified Bessel function of second type. 
When the $U(1)_\Sigma$ phase transition kicks in and the Majorana mass is suppressed by a few orders of magnitude, $Y_\mathrm{eq}$ turns into the equilibrium yield of a particle species with the smaller mass $M_f$.
The role of $Y_\mathrm{eq}$ in \cref{eq:boltzi1} is to produce DM (whereas the role of $Y$ in \cref{eq:boltzi1} is to deplete it).
What this is telling is that, although prior to the phase transition the thermal bath (represented by $Y_\mathrm{eq}$) is no longer efficiently producing DM, after the phase transition the DM has turned so much lighter that the thermal bath again has enough energy to produce it.
This means that, after the $U(1)_\Sigma$ phase transition induces DM freeze-out, there might (depending on the interaction rate) be a second period of DM production that would last until $T\approx M_{f}$.
This second phase of DM production is driven by the well known mechanism of \textit{freeze-in}.
However, a look at the right panel of \cref{fig:suppressed_decay_rate} reveals that after the $U(1)_\Sigma$ phase transition $\langle\gamma\rangle$ will be so dramatically suppressed that there is no hope of observing any efficient production by freeze-in.

\begin{figure}[bt]
\begin{center}
  \includegraphics[width=1\textwidth]{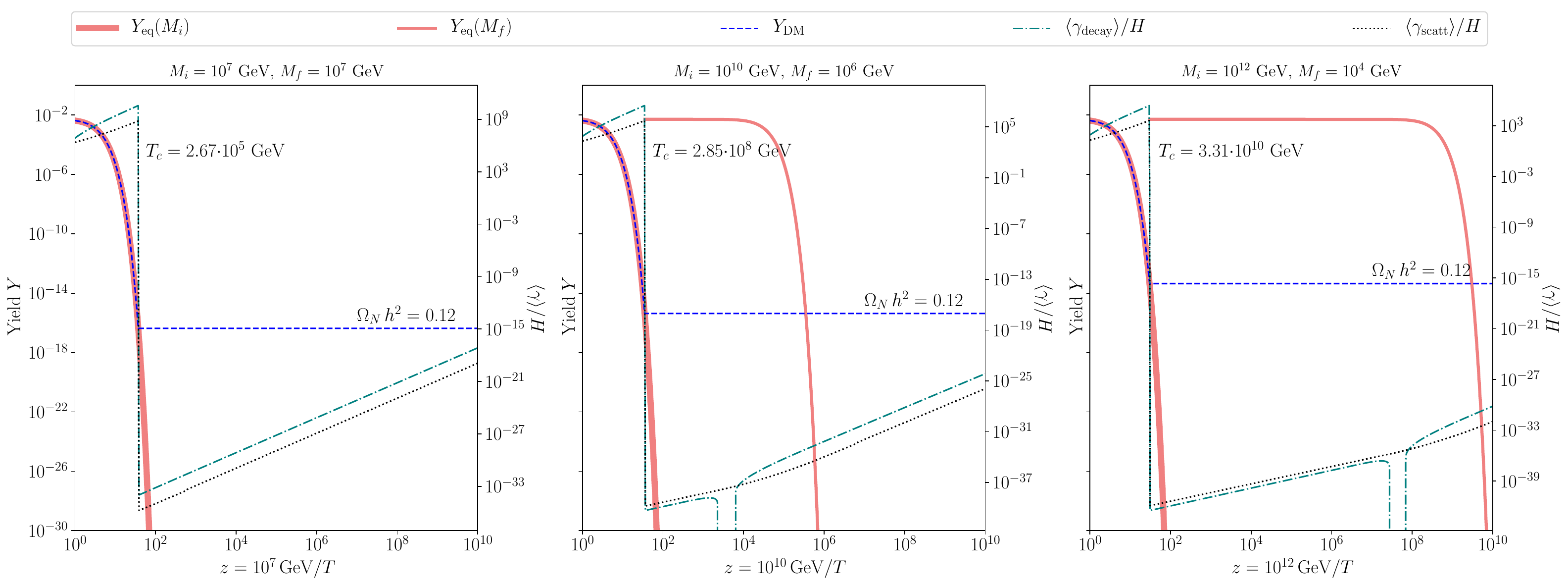}
  \caption{Solutions to the Boltzmann equation \cref{eq:boltzi1} for representative models from \cref{tab:FN-configurations}:
  M1 (\textit{left}), M2 (\textit{center}) and M3 (\textit{right}). The equilibrium yield $Y_\mathrm{eq}$ for $M_i$ and $M_f$ together with the computed yield for DM neutrinos $Y$ are plotted on the left vertical axis; all three panels share the same scale. The decay and scattering rates are plotted on the right vertical axis and span  different ranges in each panel. The critical temperature $T_c$ (or $z_c$ equivalently) was chosen such that the produced relic abundance coincides with the observed value for DM by the Planck Collaboration \cite{Aghanim:2018eyx}.}
  \label{fig:yields}
\end{center}
\end{figure}

The results of numerically solving the Boltzmann equation \eqref{eq:boltzi1} for different parameters are shown in \cref{fig:yields}.
The parameters chosen are examples representing the different models M1 (left), M2 (center) and M3 (right) from \cref{tab:FN-configurations}.
In all three cases the relic abundance of sterile neutrinos would make up $100\,\%$ of the observed dark matter of the Universe \cite{Aghanim:2018eyx}.
Notice that in the case of the M1 model, for which the FN charge of DM neutrinos is $q_N=0$, their mass is not suppressed, meaning that after the $U(1)_\Sigma$ transition they do not become relativistic again and $Y_\mathrm{eq}$ does not change.
The secondary vertical axis on the right side of each panel in \cref{fig:yields} describes the total interaction rate, which is always drastically suppressed after the $U(1)_\Sigma$ phase transition.
Indeed, because of this, models M2 and M3  do not have a second phase of DM production by freeze-in, which, as just discussed, in principle could occur.

\begin{figure}[bt]
\begin{center}
  \includegraphics[width=1\textwidth]{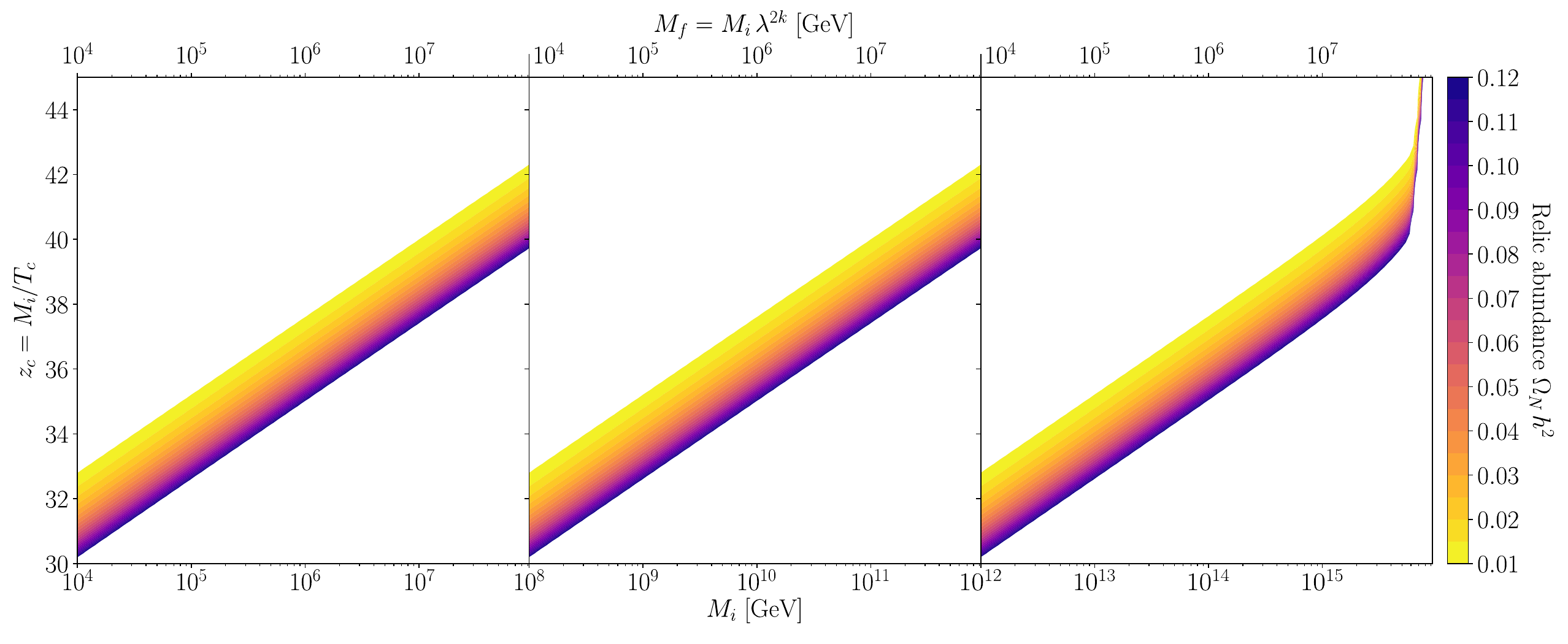}
  \caption{The regions in the parameter space where a meaningful contribution ($100\,\%$ dark shades, $10\,\%$ light shades) to the DM density of the Universe can be achieved by the framework presented in this work.}
  \label{fig:param_scan}
\end{center}
\end{figure}

Besides the DM mass, the most important parameter determining the relic abundance is the critical temperature $T_c$ at which the $U(1)_\Sigma$ phase transition occurs.
Using $z_c$ as a proxy for $T_c$, we have solved the Boltzmann equation \cref{eq:boltzi1} and computed the resulting relic abundance of DM neutrinos for models M1, M2 and M3 from \cref{tab:FN-configurations} sweeping over the complete mass range $M_i=[10^4\,\mathrm{GeV},10^{16}\,\mathrm{GeV})$.
The results are shown in \cref{fig:param_scan}, where the color scheme shows the regions in the parameter space where the relic abundance lies between $100\,\%$ (dark) and $10\,\%$ (light) of the observed value. For bare Majorana masses, anywhere between $M_i = 10^{4}\,\mathrm{GeV}$ and $M_i = 10^{16}\,\mathrm{GeV}$, the critical temperature for the $U(1)_\Sigma$ phase transition must be between $M_i/44$ and $M_i/30$ in order to produce a meaningful contribution of up to $100\,\%$ to the dark matter density of the Universe.
The upper horizontal axis shows the mass of DM neutrinos after the FN phase transition, which for each model always spreads over $[10^4\,\mathrm{GeV},10^{8}\,\mathrm{GeV})$.
This is the effective mass that the DM neutrinos would have today.
With the exception of the region around $M_i = 10^{16}\,\mathrm{GeV}$, we notice that the shape of the allowed parameter space compatible with the DM hypothesis is almost identical for all three model classes.
This has to do with the fact that although each model class has different $M_i$ ranges, the $M_f$ range is the same for all model classes, e.g.\ for a set of models M1 with $M_i = 10^{6}\,\mathrm{GeV}$, M2 with $M_i = 10^{10}\,\mathrm{GeV}$ and M3 with $M_i = 10^{14}\,\mathrm{GeV}$ the FN charges in \cref{tab:FN-configurations} and \cref{eq:initial_final_M}, result in $M_f = 10^{6}\,\mathrm{GeV}$ in all three cases.
Furthermore, if they all freeze-out at $z_c$, then they will all have the same asymptotic yield $Y_\infty$.
The relic abundance is then calculated by
\begin{align}
\Omega_N \,h^2 = \frac{Y_\infty \, s_0\,M_f}{\rho_\mathrm{crit}},
\end{align}
where $s_0$ is the entropy density today and $\rho_\mathrm{crit}$ is critical energy density of the Universe.
Thus, it is clear that if the asymptotic yield $Y_\infty$ and $M_f$ are the same for different models with different initial Majorana masses, the relic abundance will be the same.
The peculiar behaviour approaching $M_i = 10^{16}\,\mathrm{GeV}$ has to do with the fact that, for this very large bare Majorana mass, the value of $\langle \gamma_\mathrm{tot}\rangle/H$ is smaller than $1$ even at $z \sim 1$, so that even before the $U(1)_\Sigma$ transition occurs, the sterile neutrinos are not quite in thermal equilibrium and do not follow $Y_\mathrm{eq}$ so closely. 

We finally note that, if no other contribution to the neutrino mass exists, the tiny value of the Yukawa coupling $y_{\rm eff}$ after the $U(1)_\Sigma$ phase transition means that the DM right-handed neutrino does not contribute to the generation of active neutrino masses, thereby implying an essentially vanishing smallest neutrino mass. Moreover, the requirement of heavy neutrino masses above $10\,$TeV, and the small mixing after the phase transition, implies that direct detection searches will be unsuccessful. 

\section{Conclusions}\label{sec:conclusio}

The type I seesaw mechanism offers a very simple and attractive explanation for the origin and tiny size of the masses of active neutrinos and a possible solution to the problem of baryogenesis through the mechanism of leptogenesis. 
Similarly, the Froggatt-Nielsen mechanism solves the problem of flavour hierarchy in a very appealing manner.

In this work we have shown that, combining both aforementioned mechanisms, it is possible to formulate slightly extended models where the problem of the dark matter of the Universe can be addressed.
The particle content of the effective theory is extended by the three right-handed neutrinos of the seesaw mechanism, one of which is the Dark Matter particle, the Froggatt-Nielsen flavon, and the new scalar $\Sigma$. 
When the $\Sigma$ field gets its vev, the Majorana mass and Yukawa coupling of the sterile neutrinos are suppressed by their Froggatt-Nielsen charges.
This inevitably induces the freeze-out of the dark matter sterile neutrino.
Thanks to the strongly suppressed Yukawa coupling after the phase transition, the cosmic-scale-longevity of the dark matter neutrinos is guaranteed.

This result is not only valid for the specific case of an extended seesaw-Froggatt-Nielsen model, such as the one studied in this work. In principle it is actually valid much more generally, for any model where a BSM particle couples strongly to SM bath in the early Universe, and whose coupling is later suppressed, by whatever mechanism, such that the BSM particle becomes effectively decoupled. Such frameworks also could explain negative results of direct detection experiments. 

The concrete framework discussed here allows to make sterile neutrinos dark matter particles generated by freeze-out from thermal equilibrium, and works over a wide range of masses.
Moreover, additional right-handed neutrinos, with FN charges different than the DM neutrino, could be responsible for generating the active neutrino masses and also be involved in leptogenesis. These possibilities are very much worth further research. 
\\

\begin{center}
{\bf Acknowledgements}
\end{center}
WR is supported by the DFG with grant RO 2516/7-1 in the Heisenberg program.

\begin{center}
{\bf Statement of Provenance}
\end{center}
This is the Accepted Manuscript version of an article accepted for publication in \textit{Journal of Cosmology and Astroparticle Physics}. Neither SISSA Medialab Srl nor IOP Publishing Ltd. is responsible for any errors or omissions in this version of the manuscript or any version derived from it.  The Version of Record is available online at \url{https://doi.org/10.1088/1475-7516/2021/04/023}.

\bibliographystyle{unsrturl}


\end{document}